\documentclass{elsart}

\usepackage{graphics}
\usepackage{amssymb}
\usepackage{color}

\begin{document}

\begin{frontmatter}



\title{Track Reconstruction and Performance of DRIFT Directional Dark Matter Detectors using Alpha Particles}

\author[Oxy_addr]{S.Burgos},
\author[Oxy_addr]{J.Forbes},
\author[Edin_addr]{C.Ghag},
\author[UNM_addr]{M.Gold},
\author[Shef_addr]{V.A.Kudryavtsev},
\author[Shef_addr]{T.B.Lawson\corauthref{cor1}},
\ead{t.lawson@sheffield.ac.uk}
\corauth[cor1]{Tel. +44-(0)114-2223517; fax. +44-(0)114-2728079}
\author[UNM_addr]{D.Loomba},
\author[Shef_addr]{P.Majewski},
\author[Shef_addr]{D.Muna},
\author[Edin_addr]{A.StJ.Murphy},
\author[Shef_addr]{G.G.Nicklin},
\author[Shef_addr]{S.M.Paling},
\author[Oxy_addr]{A.Petkov},
\author[Edin_addr]{S.J.S.Plank},
\author[Shef_addr]{M.Robinson},
\author[UNM_addr]{N.Sanghi},
\author[RAL_addr]{N.J.T.Smith},
\author[Oxy_addr]{D.P.Snowden-Ifft},
\author[Shef_addr]{N.J.C.Spooner},
\author[Imperial_addr]{T.J.Sumner}
\author[UNM_addr]{J.Turk},
\author[Shef_addr]{E.Tziaferi},
\address[Oxy_addr]{Department of Physics, Occidental College, Los Angeles, CA 90041, USA}
\address[Shef_addr]{Department of Physics and Astronomy, University of Sheffield, S3 7RH, UK}
\address[Edin_addr]{School of Physics, University of Edinburgh, EH9 3JZ, UK}
\address[UNM_addr]{Department of Physics and Astronomy, University of New Mexico, NM 87131, USA}
\address[RAL_addr]{Particle Physics Department, CCLRC Rutherford Appleton Laboratory, Chilton, Didcot, Oxon OX11 0QX, UK}
\address[Imperial_addr]{Blackett Laboratory, Imperial College of Science, Technology and Medicine, London SW7 2BZ, UK}
\begin{abstract}
First results are presented from an analysis of data from the DRIFT-IIa and DRIFT-IIb directional dark matter detectors at Boulby Mine in which alpha particle tracks were reconstructed and used to characterise detector performance---an important step towards optimising directional technology. The drift velocity in DRIFT-IIa was $59.3 \pm 0.2 \mbox{ (stat)} \pm 7.5 \mbox{ (sys)}$ ms$^{-1}$ based on an analysis of naturally-occurring alpha-emitting background. The drift velocity in DRIFT-IIb was $57 \pm 1\mbox{ (stat)} \pm 3 \mbox{ (sys)}$ ms$^{-1}$ determined by the analysis of alpha particle tracks from a $^{210}$Po source. 3D range reconstruction and energy spectra were used to identify alpha particles from the decay of $^{222}$Rn, $^{218}$Po, $^{220}$Rn and $^{216}$Po. This study found that $(22 \pm 2)$\% of $^{218}$Po progeny (from $^{222}$Rn decay) are produced with no net charge in 40 Torr CS$_2$. For $^{216}$Po progeny (from $^{220}$Rn decay) the uncharged fraction is $(100^{+0}_{-35})$\%.
\end{abstract}


\end{frontmatter}

% main text
\section{Introduction}
\label{sec:intro}
The DRIFT project aims to develop and operate the first underground time projection chamber (TPC) array suitable for observing and reconstructing WIMP-induced nuclear recoil tracks with enough precision to provide a signature of the local WIMP Galactic halo~\cite{Morgan_Green_Spooner05,Morgan_th,Copi_Krauss01,Bergstrom00}. The development of the DRIFT Negative Ion TPC (NITPC) detector concept is covered in~\cite{Ifft00,Martoff00,Ohnuki01,Ifft03,Lawson_thesis,D-I_tech,Lightfoot07,Lawson_tech05}. A full desription of the latest detector design, DRIFT-II, including installation and operation of the first modules, DRIFT-IIa and DRIFT-IIb, at Boulby Mine is covered in~\cite{Lawson_tech05}. The purpose of this paper is to bring together results concerning the behaviour of DRIFT-IIa and DRIFT-IIb as evidenced by observation of various classes of alpha particle events. The results in this paper are complementary to those covered in~\cite{D2science_06}, where the alpha decay of radon progeny were found to produce recoil-like events.

Although alpha particle events are, ultimately, an unwanted background in a rare event detector, their presence in DRIFT allows investigations of event discrimination, track reconstruction, directional sensitivity and other data analysis capabilities. Determination of the accuracy of ion transport codes (e.g. SRIM2003~\cite{SRIM}) is also possible\footnote{The authors are aware of a new version of this code (SRIM2006), however this `includes no changes to the basic calculations of SRIM-2003'\cite{SRIM}.}. Alpha particles of a few MeV in 40 Torr CS$_2$ produce essentially straight tracks typically 300--700 mm long~\cite{SRIM} that are easily rejected from a dark matter analysis by range and energy cuts (see \S\ref{sec:bg_alphas} and \cite{D2science_06}). Such tracks provide an ideal means for developing track reconstruction algorithms. Studies (see \S\ref{sec:alpha_types}~and~\S\ref{sec:MCs}) of the alpha particle event populations found in DRIFT-II data also provide important information to allow elimination of such background events from present and future modules.

The remainder of this paper will cover detector design (\S\ref{sec:design}), data analysis (\S\ref{sec:analysis}), a discussion of the different types of alpha particle events observed (\S\ref{sec:alpha_types}), simulations of alpha activity (\S\ref{sec:MCs}), determination of the drift velocity and range-discrimination of different alpha particle tracks (\S\ref{sec:drift_vel}).

\section{DRIFT Detector Description}
\label{sec:design}
The DRIFT project aims to make continuing improvements to the design of DRIFT detector modules. DRIFT-IIb is an upgraded version of DRIFT-IIa~\cite{Lawson_tech05} in which the significant differences are a simplified arrangement of electronics and the replacement of copper fieldrings with stainless steel versions. Briefly, each detector consists of a $1.5 \times 1.5 \times 1.5$ m$^3$ stainless steel vacuum vessel housing a 1 m$^3$ dual low pressure TPC filled with CS$_2$ vapour at 40 Torr. The two modules were operated at different mean drift fields (620 V/cm in DRIFT-IIa, 581 V/cm in DRIFT-IIb). The two 0.5 m long drift regions share a central plane of 512 stainless steel wires of 20 $\mu$m diameter spaced 2 mm apart. This arrangement forms a highly transparent drift cathode. Readout of charge deposited within the drift regions is via two 1 m$^2$ multi wire proportional chambers (MWPCs), 50 cm either side of the central cathode. The MWPCs consist of grid planes of 512 stainless steel wires of 100 $\mu$m diameter. The grid planes are spaced 1 cm either side of an anode plane comprising 512 wires of 20 $\mu$m diameter. These are arranged orthogonal to the wires in the grids. The wire spacing in all the planes is 2 mm. Veto regions are instrumented around the edges of the wire planes.

\begin{figure} % file_id = 7781, ev. 77
\begin{center}
\scalebox{0.4}{\includegraphics{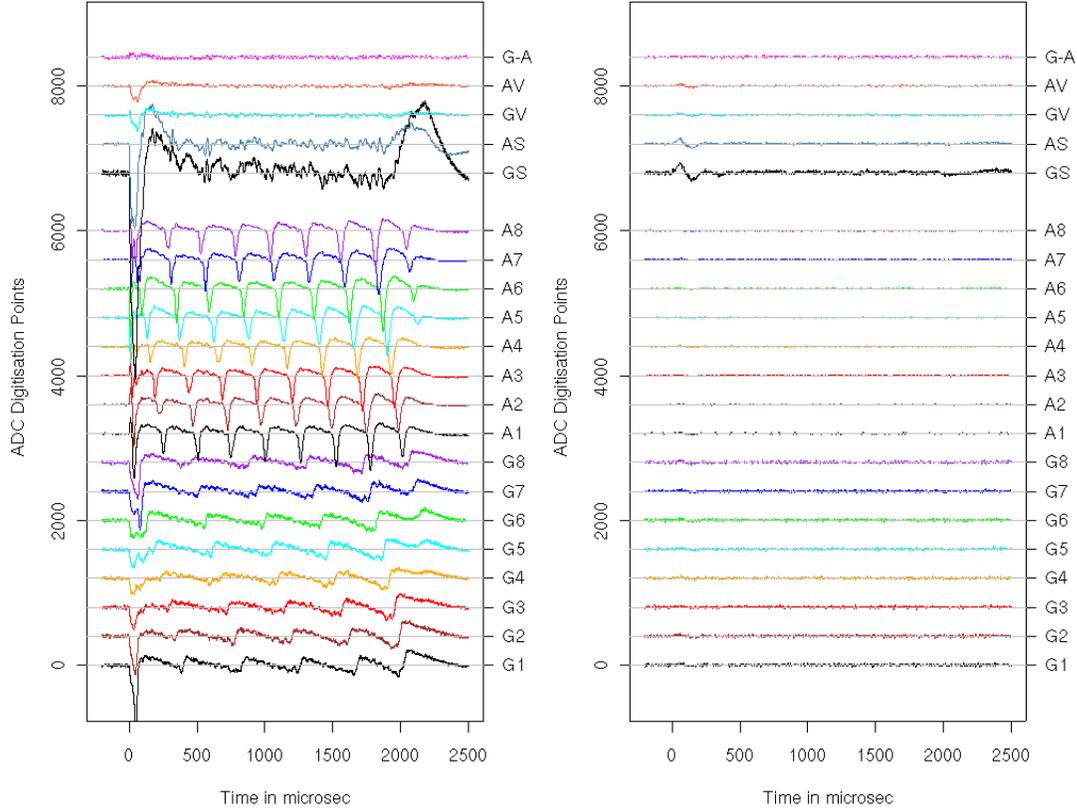}} % typ_alpha_event-greyscale.eps
\caption{An example alpha particle track. Each trace is derived from signals produced on every eighth wire in the MWPC and represents the amplified charge signal plotted over time. The event record consists of 1 ms of pretrigger plus 4 ms of postrigger data (trigger at $t=0$). Tracks that cross more than eight wires `wrap around' and reappear on the original channel. The labels down the right hand side (in order from top) refer to: difference between grid and anode vetos; anode veto; grid veto; sum of anode channels; sum of grid channels; anode channels ($\times8$); grid channels ($\times8$).\label{fig:typ_alpha_event}}
\end{center}
\end{figure}

The central section of the anode and inner grid planes are each grouped down to eight outputs. As such, any track that extends a distance greater than that sampled by eight wires ($>16$ mm) will reappear on the same output channnels at a later time in the event record (see~\cite{Lawson_tech05} for details of DRIFT data format). This phenomenon was characteristic of all the alpha particle tracks observed. Since high energy alpha particles have a range greater than 300 mm in 40 Torr CS$_2$, the hit pattern on the eight outputs repeats many times. Figure~\ref{fig:typ_alpha_event} depicts waveforms for an example alpha particle event.

\section{Analysis Procedure}
\label{sec:analysis}
\subsection{Data Reduction and Parameterisation}
\label{sec:reduction}
A dedicated analysis was developed to specifically select alpha particle like events from the data. Alpha particle tracks are sufficiently different from those of nuclear recoils that the analysis procedures described in Ref~\cite{D2science_06} would not have been appropriate.

During data acquisition all electronics channels\footnote{A total of 36 channels comprising eight anodes, eight grids, a grid veto and an anode veto from each of the two MWPCs.} had their signals digitised and recorded when any anode signal exceeded the hardware threshold (200 ADC points, where 2048 ADCs is equivalent to one Volt). Throughout this paper `above threshold' means that the absolute value of the waveform exceeded the absolute value of the threshold unless specified otherwise. Once acquired, three virtual waveforms for each MWPC (the grid sum, the anode sum and the veto difference---see Figure~\ref{fig:typ_alpha_event}) were then constructed from existing waveforms. All waveforms (real and virtual) then underwent a fast Fourier transform (FFT) to identify frequency components introduced by unwanted signal pickup (specifically, components around 50 kHz associated with the drift cathode power supply). The unwanted frequencies were notched out and the result inverse Fourier transformed. For each waveform, the baseline was calculated using the mean signal value in the pretrigger region between $-900$ and $-100 \mu$s from the trigger time (a total of 1 ms is recorded prior to the trigger). The data were then smoothed using a Savitzky-Golay filter~\cite{NumRepC} using a 4th-degree polynomial. This allowed the effective analysis threshold to be reduced to $\sim 9$ ADC points.

The raw and smoothed data were reduced to a number of parameters that were stored in several tables within a relational database that could then by queried by subsequent analysis routines. The following is a brief description of the key parameters used. The names of all analysis parameters in this paper are refered to using {\it italics}.

\subsection{Analysis Parameters}
For each output channel, the voltage waveforms were scanned for pulses, i.e. sections which exceeded threshold. These sections are refered to as signal profiles. A signal profile was deemed to start and end at the nearest points, either side of the peak, where the waveform crossed the baseline. The main parameters extracted from the raw data are summarised in Table~\ref{tab:an_params}.

\begin{table}
\caption{The main analysis parameters extracted from the raw data.\label{tab:an_params}}
\begin{center}
\begin{tabular}{lp{95mm}}
{\it polarity:} & $+1$ for positive-going signal-profiles, $-1$ for negative.\\
{\it height:} & Maximum absolute excursion of waveform while above threshold (in ADCs).\\
{\it voltage\_weighted\_time:} & `Centre-of-gravity' in time, of a signal-profile.\\
{\it anode\_area:} & Integrated area of all {\it polarity} $=-1$ anode signal-profiles.\\
{\it wires:} & Number of {\it polarity} $=-1$ anode signal-profiles with {\it height} $>45$ ADCs, equivalent to the number of physical anode wires in the detector on which significant charge was deposited.\\
{\it start\_time:} & Start of the FWHM of the first {\it polarity} $=-1$ anode signal-profile in a track.\\
{\it end\_time:} & End of the FWHM of the last {\it polarity} $=-1$ anode signal-profile in a track.\\
\end{tabular}
\end{center}
\end{table}

Individual tracks were identified by grouping together negative polarity anode signal profiles in time order to form one `track' entity. This algorithm relied on the assumption that tracks were well approximated by a straight line. This is a reasonable assumption for high energy alpha particle tracks. In a given event, tracks on each MWPC were recorded separately. Thus, a particle track straddling the central cathode and depositing charge in each side of the detector was counted as two separate tracks. These partial tracks were recombined at a later stage of the analysis.

The high charge-density of alpha particle tracks resulted in large pulses in the anode output waveforms. Thus a high threshold (45 ADC) was chosen to select these without contamination from sparks, nuclear recoils and gamma ray events. It was found, from a random selection of tracks, that reducing the pulse height threshold from 45 ADCs to 30 ADCs increased the mean $x$- and $z$-components by less than 0.2\%. Increasing the threshold to 50 ADCs resulted in the loss of $5 \pm 2$\% of alpha particle tracks with the $x$- and $z$-components of the remaining tracks left unnaffected.

The $x$ component of a track (parallel to the wire plane, perpendicular to anodes), was calculated using the relation $\Delta x= (\mbox{\it wires}-1)\times 2$ mm. Grid signals were smaller than those on the anodes, requiring a different method for obtaining the track $y$ component (parallel to the wire plane, perpendicular to grids): $\Delta y= (\Delta t / \overline{dt} - 1)\times 2$ mm, where $\Delta t$ is the difference between the {\it start\_time} and {\it end\_time} and $\overline{dt}$ is the mean time between successive negative polarity {\it voltage\_weighted\_time}s in the grid waveforms occuring between the {\it start\_time} and {\it end\_time}. The threshold for accepting signal profiles on the grids was 9 ADCs.

\subsection{Track Definitions}
\label{sec:flags}
Tracks having {\it wires} $> 8$, distributed across all 8 anode channels (i.e. at least one signal profile per channel) were flagged in the database as candidate alpha particle tracks. A major part of the analysis involved the selection of alpha particle events that crossed the cathode from one side of the detector to the other without crossing a veto region. Such events were labelled `Gold Plated Cathode Crossers' (GPCCs) and were selected with the following additional cuts:
\begin{enumerate}
\item[1.]{An alpha particle track, as defined above, appeared on both MWPCs,}
\item[2.]{The event had no pulses on the veto difference line with {\it height} $> 45$ ADCs.}
\item[3.]{$\mbox{\it anode\_area}/\mbox{\it wires}< 30$ V$\times\mu$s. This cut eliminated occasional amplifier feedback---a problem that affected the highest amplitude events and was resolved in DRIFT-IIb with replacement preamplifiers. This cut alone retained $>87\%$ of all candidate alpha tracks.}
\item[4.]{{\it anode\_area} $> 30$ V$\times\mu$s. This eliminated events such as sparks that deposited charge on all readouts but had a combined charge less than that expected for alpha particle events. This cut alone retained greater than 98\% of candidate alpha tracks.}
\item[5.]{{\it start\_time} greater than -100 $\mu$s. This cut alone accepted greater than 96\% of events and was introduced to remove events in which the pretrigger region contained pulses from the preceeding trigger.}
\item[6.]{{\it end\_time} $<3000\ \mu$s. This provided a 1000 $\mu$s buffer between the track {\it end\_time} and the end of the record, ensuring that only tracks with a mean {\it voltage\_weighted\_time} separation $> 1000\ \mu$s, on a given anode, could evade this cut illegally (equivalent to $<0.1\%$ of all candidate alpha particle tracks).}
\end{enumerate}

In summary, none of the cuts mentioned above had a significant effect on the measured track range or derived drift velocity (see \S\ref{sec:drift_vel}), since the dimensions of the tracks were not affected. However, cut 6 did constrain the orientations of the longest tracks in DRIFT-IIa, accepted as GPCCs, to be greater than $\sim 70^{\circ}$ from the normal to the wire planes.

\section{Types of Alpha Particle Events}
\label{sec:alpha_types}
\subsection{Alpha Particle Event Parameters}
Alpha particles were expected in the detector due to decay of $^{222}$Rn ($\tau_{1/2} = 3.8$ days, decay energy = 5.59 MeV) and subsequent alpha decays of $^{218}$Po and $^{214}$Po. Radon progeny are usually produced as positively charged ions in CS$_2$ and so are attracted to the drift cathode where they plate out. Alpha decays from $^{210}$Po are ignored in this analysis since these occur below the long-lived $^{210}$Pb ($\tau_{1/2} = 22.2$ years) in the $^{238}$U decay chain and subsequent event rates are insignificant.

In many materials traces of $^{232}$Th are also present at comparable levels to those of $^{238}$U~\cite{UKDM_rad_data}. This contamination gives rise to the much shorter-lived $^{220}$Rn ($E_{\alpha} = 6.288$ MeV, $\tau_{1/2} = 55.6$ s). The daughter nucleus from this decay, $^{216}$Po, is also produced positively charged in CS$_2$ and although its half life is relatively short ($\tau_{1/2} = 145$ ms) this is plenty of time for it to be drifted to the central cathode and be plated out on a cathode wire surface. There are two subsequent alpha decays (from $^{216}$Po and $^{212}$Po) plus two $\beta$ decays with the decay sequence ending in stable $^{208}$Pb.

\begin{figure}
\begin{center}
\scalebox{0.55}{\includegraphics{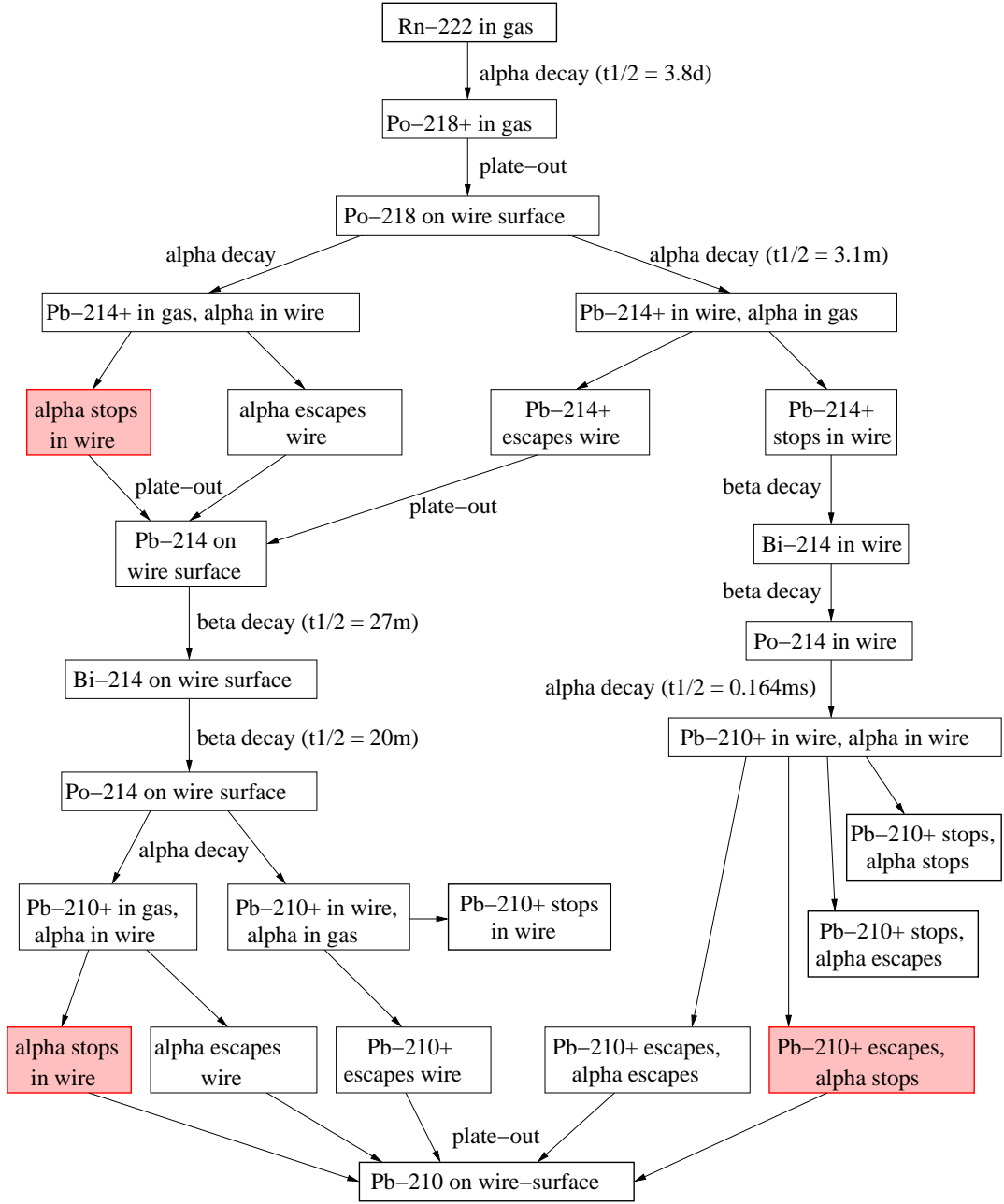}}
\caption{Flow diagram of one decay cycle, starting with the alpha-decay of $^{222}$Rn and ending with the long-lived ($\tau_{1/2} \sim 22$ years) $^{210}$Pb. The three shaded boxes denote events that produce a recoil with no accompanying alpha track observable. These events are termed radon progeny recoils (RPRs) and are an important background since they mimic the recoil signature expected for wimp-nucleon interactions.\label{fig:222decay}}
\end{center}
\end{figure}

\begin{figure}
\begin{center}
\scalebox{0.55}{\includegraphics{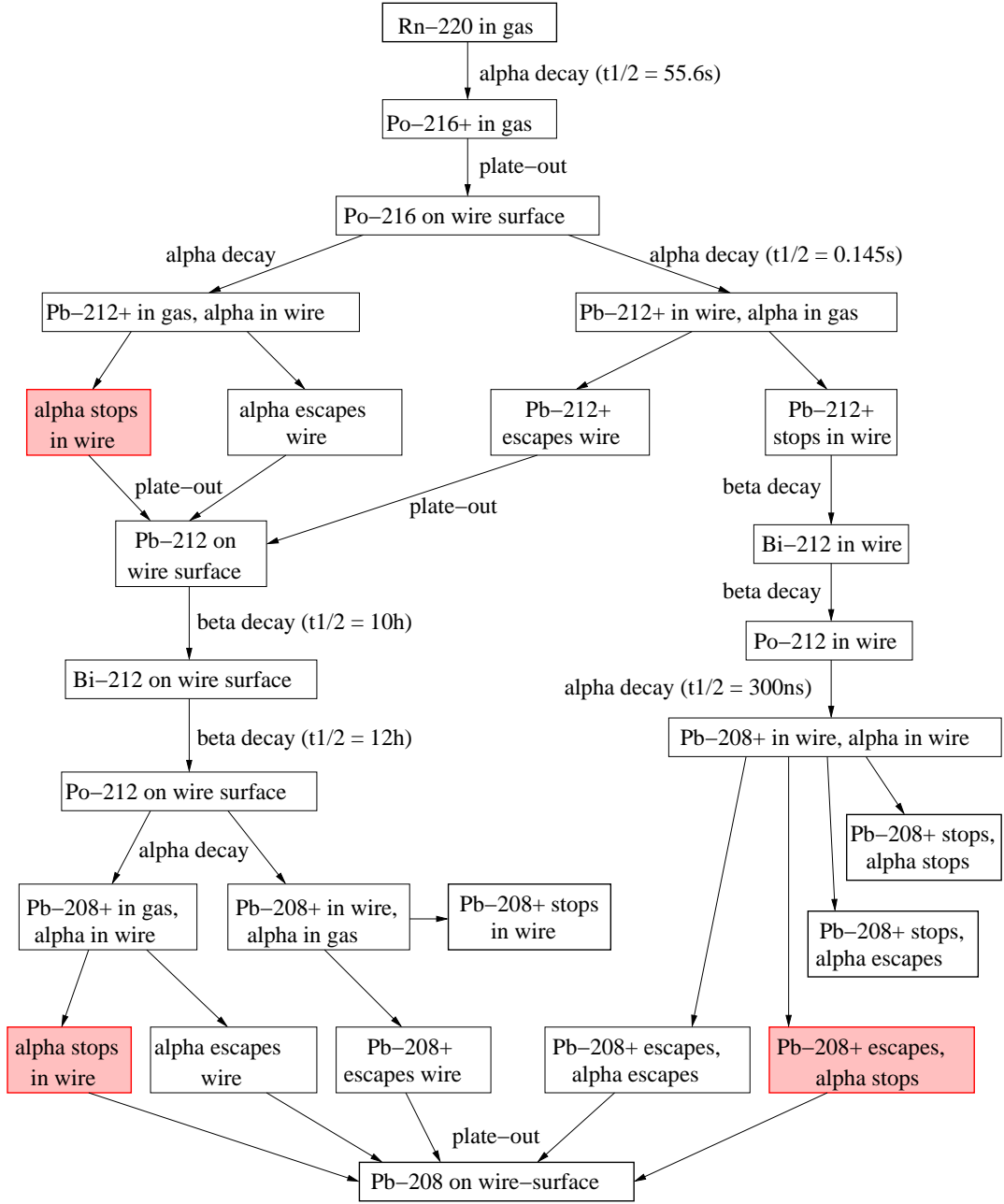}}
\caption{Flow diagram, analagous to that depicted in Figure~\ref{fig:222decay}, starting with the alpha-decay of $^{220}$Rn and ending with stable $^{208}$Pb.\label{fig:220decay}}
\end{center}
\end{figure}

A key issue in the decay sequence is the generation of recoiling progeny nuclei. Figures~\ref{fig:222decay}~and~\ref{fig:220decay} illustrate the decay sequence from $^{222}$Rn to $^{210}$Pb and $^{220}$Rn to $^{208}$Pb, respectively, showing how interaction with a single cathode wire can allow low energy recoiling progeny to be observed without detection of the accompanying alpha particle, mimicking WIMP-nucleon interactions in the detector.


From arguments and calculations presented in \S\ref{sec:drift_vel}, the drift velocities in the DRIFT-IIa and DRIFT-IIb detectors are determined to be $59.3 \pm 0.2\mbox{ (stat)} \pm 7.5\mbox{ (sys)}$ ms$^{-1}$ and $56 \pm 1.7\mbox{ (stat)} \pm 3\mbox{ (sys)}$ ms$^{-1}$, respectively. A condition imposed by the analysis of alpha particle events required that ionization from an event completed its deposition in either MWPC within a time period of $t = 3$ ms. Combined with the drift velocity, this limits the z-component of any track recorded by an individual MWPC to be $z_{max} < 178$ mm (DRIFT-IIa) or $< 168$ mm (DRIFT-IIb). Modeling alpha particle tracks in 40 Torr CS$_2$ using SRIM-2003~\cite{SRIM} indicates typical ranges of 300--700 mm. These ranges are listed in Table~\ref{tab:alpha_ranges}. Since these ranges all exceed $z_{max}$, only a subset of alpha particle tracks are accepted by the analysis. The fraction of events that are accepted is covered in \S\ref{sec:MCs}. Tracks on a given MWPC that extend further in z than $z_{max}$ may still have been unvetoed but this fact could not be determined from the data. Such events that `fell off' the end of the data record were also incomplete and so could not contribute to energy spectra. The effects of this and other cuts are discussed in \S\ref{sec:MCs}.
\begin{table}
\begin{center}
\caption{Energies and ranges of alpha particle tracks and resultant recoiling progeny expected in 40 Torr CS$_2$ due to decay of $^{222}$Rn and $^{220}$Rn and their respective progeny. Ranges and longitudinal straggling (`Long. strag.') are as simulated by SRIM2003\cite{SRIM}. Note that the polonium decays will typically occur on or near the surface of a cathode wire so, for example, an outgoing alpha particle will implant the resulting recoil within the wire and vice versa.\label{tab:alpha_ranges}}
\vspace{2mm}
\begin{tabular}{r|ccc||r|ccc}
Isotope & $E_\alpha$ & Range & Long.  & Recoil & $E_{\mbox{recoil}}$ & Range    & Long.\\
        & (MeV)      & (mm)  & strag. &        & (keV)               & ($\mu$m) & strag.\\
        &            &       & (mm)   &        &                     &          & ($\mu$m)\\
\hline
$^{222}$Rn & 5.48948 & 334 & 13.4 & $^{218}$Po & 100.82              & 577.91   & 119.61\\
$^{218}$Po & 6.00235 & 383 & 15.3 & $^{214}$Pb & 112.33              & 628.54   & 129.62\\
$^{214}$Po & 7.68682 & 567 & 23.3 & $^{210}$Pb & 146.64              & 745.25   & 149.99\\
\hline
$^{220}$Rn & 6.288   & 413 & 16.3 & $^{216}$Po & 116.5               & 631.99   & 129.08\\
$^{216}$Po & 6.778   & 464 & 18.1 & $^{212}$Pb & 127.9               & 682.27   & 139.07\\
$^{212}$Po & 8.785   & 701 & 30.0 & $^{208}$Pb & 168.9               & 818.14   & 162.45\\
\end{tabular}
\end{center}
\end{table}

\subsection{Alpha Particles from Decay of $^{222}$Rn, $^{220}$Rn and uncharged Radon Progeny}
\begin{figure}% ev76_20050620-01-0078
\begin{center}
\scalebox{0.49}{\includegraphics{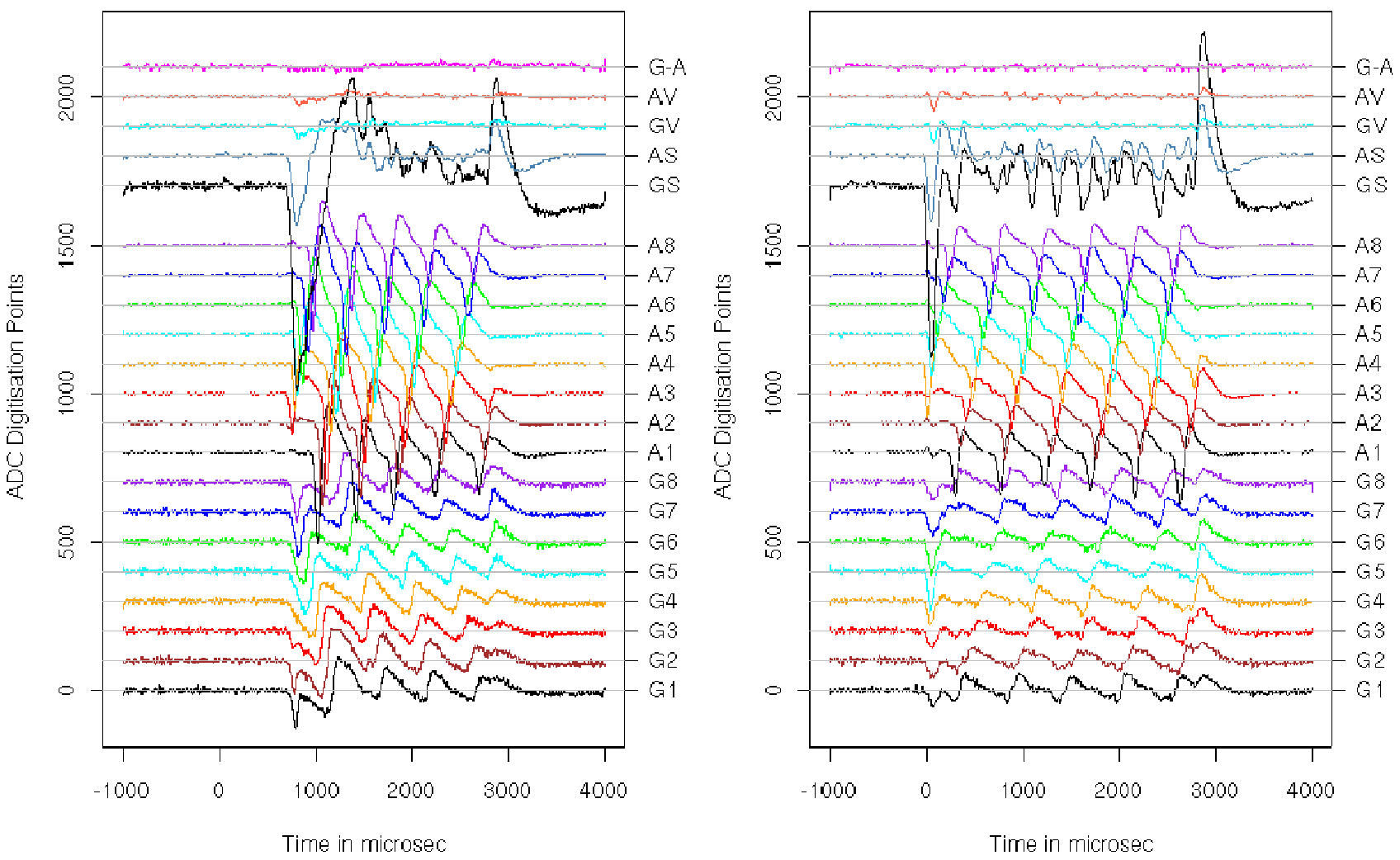}}% cc_alpha-greyscale.eps
\caption{A cathode-crossing alpha particle track due to radon decay (a GPCC event).\label{fig:cath_crosser}}
\end{center}
\end{figure}

Because $^{222}$Rn, $^{220}$Rn and their uncharged progeny are the only effectively gaseous alpha sources under consideration this is the only case in which an alpha particle track could both start and end in the gas, under the assumption that dust contamination in the gas fill is negligible. Thus only these types of events could produce tracks that cross the cathode and satisfy the other GPCC criteria (see \S\ref{sec:flags} and Figure~\ref{fig:cath_crosser}).

\begin{figure}
\begin{center}
\scalebox{0.7}{\includegraphics{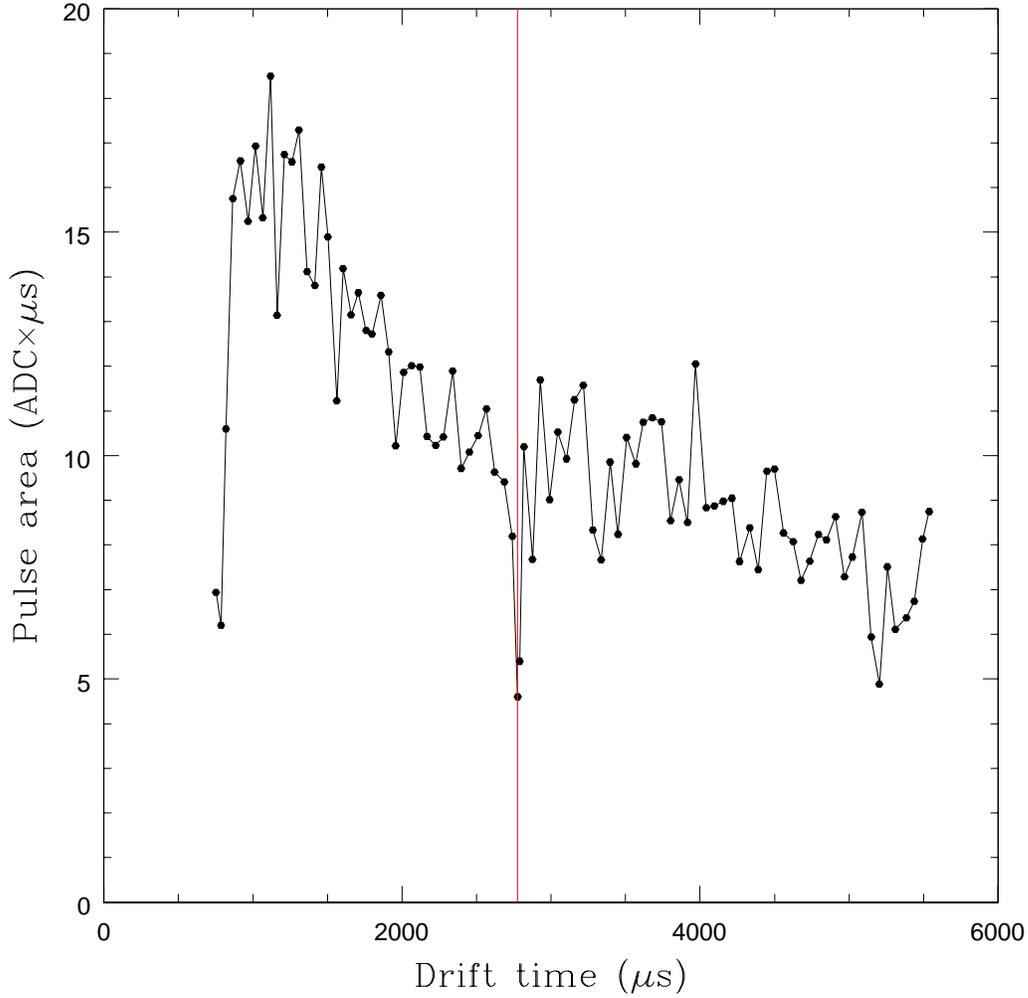}}
\caption{Plot of anode pulse area against drift time (proportional to extent in $z$-direction) for the alpha particle event shown in Figure~\ref{fig:cath_crosser}. The time-sequence of the pulses on the right detector has been reversed and added to the end of the left-hand pulse sequence to allow the whole track to be plotted. The Bragg peak is clearly visible on the left. The vertical line marks the position of the central cathode.\label{fig:Bragg_curve}}
\end{center}
\end{figure}

\begin{figure}
\begin{center}
\scalebox{0.7}{\includegraphics{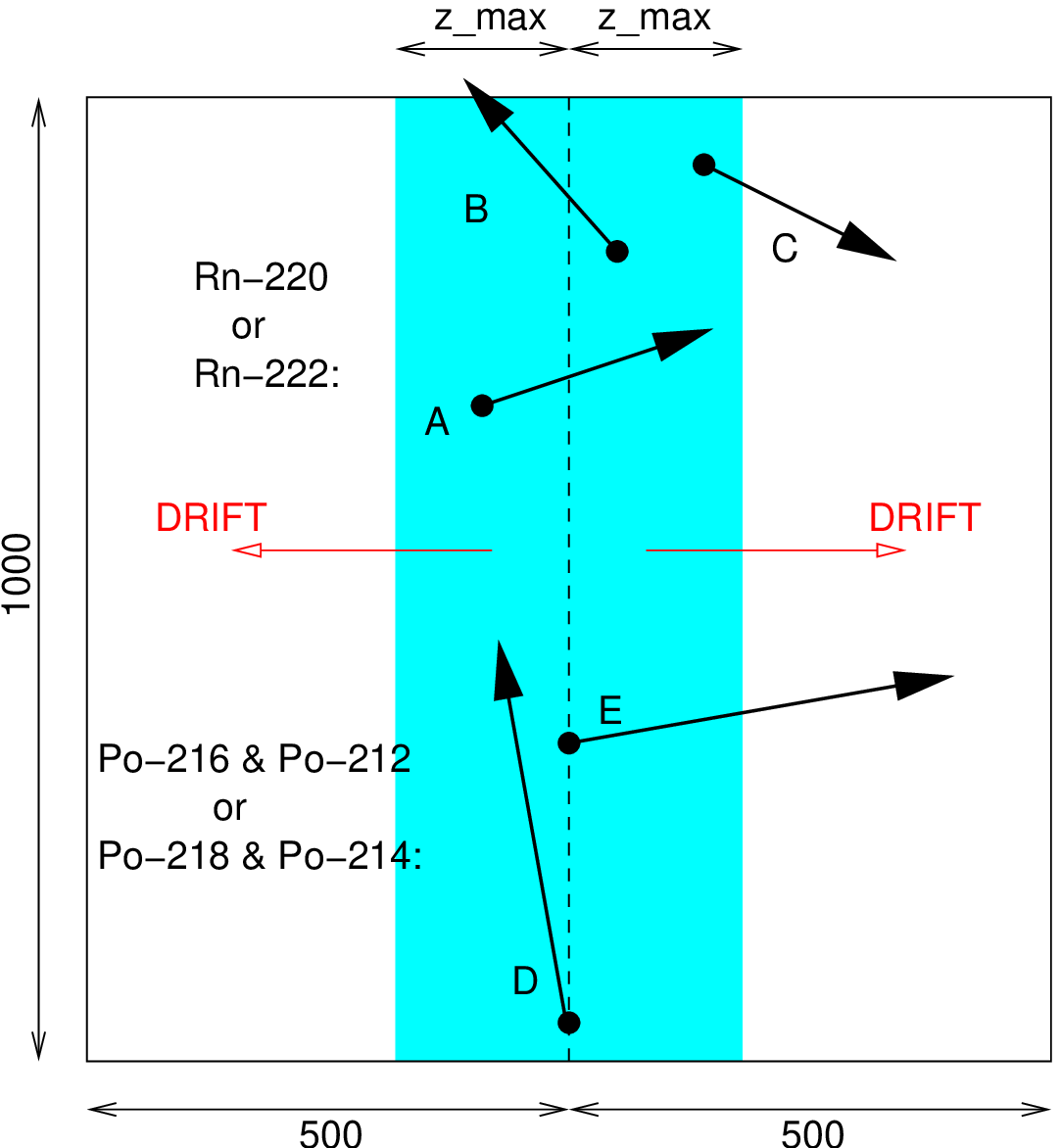}}
\caption{Illustration of different classes of alpha particle event. The 1 m$^3$ fiducial volume is shown bisected by the central cathode (dimensions in mm). Tracks that cross the cathode but remain inside the fiducial volume and have z components $< z_{max}$ (A) are counted as `gold-plated cathode crossers' (GPCCs, see main text for definition). These are used as a proportionate measure of the total radon event population (A, B and C, where B and C represent radon decay events that fail one or more GPCC selection criteria). Charged polonium decays (D and E) never cross the cathode, since, typically, they plate out on a cathode wire before decaying. Note that the higher range of polonium decays means that a smaller proportion of their total population would be accepted by the $z_{max}$ cut.\label{fig:drift_layout}}
\end{center}
\end{figure}

Radon and uncharged progeny events were characterised by (i) a clear, extended track appearing simultaneously on both sides of the detector with an extent in the z direction less than $z_{max}$; (ii) a lack of significant veto signals and (iii) simultaneous end times for tracks on each MWPC, corresponding with the point where the track crossed the cathode plane. In all cases, the changing mean peak height along the track follows the variation in dE/dx, producing a characteristic Bragg curve that reveals the direction of the original alpha particle. Figure~\ref{fig:Bragg_curve} shows a representation of the Bragg curve for the track in Figure~\ref{fig:cath_crosser}. The Bragg peak occurs in the left detector in this case indicating that the original decay occured on the right. The position of the central cathode is also shown, coinciding with a slight suppression of pulse area due to drift-field distortions around the cathode wires and charge division between the two sides of the detector. Figure~\ref{fig:drift_layout} illustrates the positioning of a typical cathode crosser track within the detector, along with other classes of alpha particle event.

\subsection{Alpha Particles from Decay of $^{218}$Po, $^{214}$Po, $^{216}$Po and $^{212}$Po}
Decay of $^{222}$Rn produces $^{218}$Po in the form of a positive ion approximately 85\% of the time in air~\cite{nazaroff_nero88}. Once produced, the uncharged $^{218}$Po fraction will be unaffected by the drift field and will thermally diffuse within the gas until it decays. The charged $^{218}$Po recoils, meanwhile, would be expected to quickly plate out onto the cathode wires~\cite{McLaughlin99}. One can expect subsequent alpha particles in the decay chain to have their origins on the cathode plane rather than in the gas. Some of the resulting tracks will stop in the gas without entering a veto region. This set of criteria could also be met by alpha particle tracks from radon decay that do not cross the cathode. An identical process occurs in the case of $^{220}$Rn progeny. Examples of this class of alpha particle track are shown in Figure~\ref{fig:drift_layout} (C, D and E).

\section{Drift Velocity Measurements}
\label{sec:drift_vel}
Knowledge of the anion drift velocity in the fiducial regions of the DRIFT-II detectors was crucial for accurate track reconstruction. To this end populations of alpha particles were observed and the drift velocity was measured via two techniques outlined in \S\ref{sec:Po210_alphas} and \S\ref{sec:bg_alphas} below.

\subsection{Field Simulations}
\label{sec:sims}
The drift velocity $v_d$ of CS$_2^-$ ions is related to the mean drift-field $E$ by $v_d = \mu E$ where $\mu$ is the ion mobility. Previous studies~\cite{Martoff00} using mixtures of CS$_2$ with Xe or Ar/CH$_4$ demonstrated this linearity, however, other studies of ion diffusion in pure CS$_2$~\cite{Ohnuki01} revealed a slight deviation from the expected linear relationship. Both these studies used considerably lower drift fields than those employed in this work and so it is helpful to extend our understanding to cover the regime in which DRIFT-II is currently operating. 

In order to fully understand the drift of charge into the MWPCs and subsequent signal development, detailed simulations of the electrostatics and avalanche process have been carried out. The results of this work will be covered in full in an upcoming paper~\cite{majewski07}. In summary, the electric field was calculated in 3D using MAXWELL~\cite{ansoft} finite element software in combination with the GARFIELD~\cite{garfield} gaseous detector code. The purpose of these calculations was to allow direct comparisons of drift velocity between different modules operated with different mean drift fields. The simulations confirmed a uniform drift field in both detector modules. The drift field in DRIFT-IIa was found to be 620 V/cm. In DRIFT-IIb the corresponding value was 581 V/cm. A simple linear dependance of drift velocity on drift field was assumed.

\subsection{Collimated Alpha Particles}
\label{sec:Po210_alphas}
\begin{figure}
\begin{center}
\scalebox{0.5}{\includegraphics{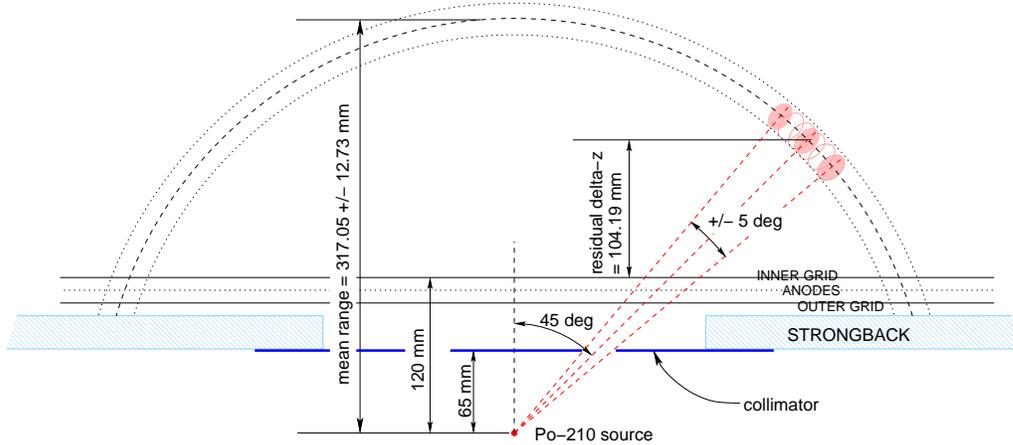}}
\caption{Setup for measuring drift velocity. A $^{210}$Po source was positioned as shown. 5.304 MeV alpha particles from this source were collimated with a $250\ \mu$m thick polyester film with a nearly complete annulus cut out of it such that the path of the alpha particles subtended an angle of $(45 \pm 5)^{\circ}$ to the normal to the wire planes. The mean residual range in the drift ($z$) direction from the inner grid was then 104.19 mm, assuming the range of tracks was as predicted by SRIM2003. The mean drift velocity $\overline{v_D}$ was then deduced from the mean duration $\overline{\Delta t}$ over which the ionization from these events produced signals on the anodes.\label{fig:v-drift_from_alphas}}
\end{center}
\end{figure}

Alpha particles from a thin-film $^{210}$Po source (alpha particle energy 5.304 MeV) were directed into the fiducial volume of the DRIFT-IIb detector as shown in Figure~\ref{fig:v-drift_from_alphas}. Alpha particles from the source with range $R$ mm were collimated to an angle of $(45 \pm 5) ^{\circ}$ by a $250\ \mu$m thick polyester mask attached to the back of the MWPC strongback (see Figure~\ref{fig:v-drift_from_alphas}). On average the alpha particles stopped $\overline{\Delta z}$ mm above the inner grid. By measuring the mean time $\overline{\Delta t}$ over which these alpha particle tracks deposited charge on the anode (grid to anode drift times are assumed to be small) the drift velocity was calculated.

The distribution of alpha particle tracks within the $x$-$y$ plane was investigated to ensure that reconstructed tracks matched the expected distribution. A concern was that the small size of the grid signals in early DRIFT-IIb operation would result in an underestimate of the track $y$-component $\Delta y$, since some signal-profiles would fall below the grid threshold and so not contribute to the measurement. This effect was found to reduce the apparent $\Delta y$ measurement by approximately a factor of two. Since this implied a sytematic error $\sigma_{\Delta y} \sim \Delta y$, events were selected in which the apparent $y$-component was less than 25 mm and the $x$-component was greater than 100 mm, to reduce the influence of this error on tracks reconstructed in the $x$-$y$ plane. As such, the total number of events was reduced from 1410 to 241.

\begin{figure}
\begin{center}
\scalebox{0.7}{\includegraphics{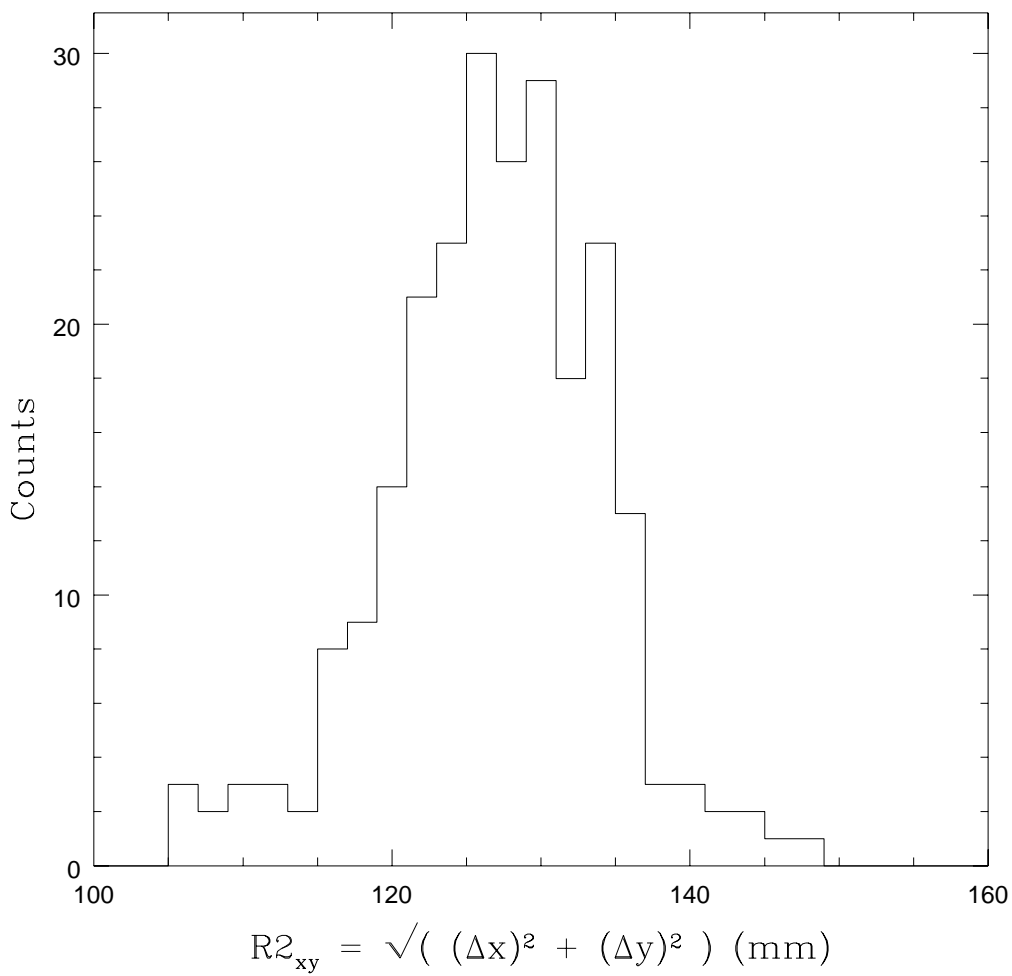}}
\caption{Distribution of the 2-dimensional range $R2_{xy}$ calculated using Equation~\ref{eq:R2xy-def} for collimated $^{210}$Po alpha particles in DRIFT-IIb. Only tracks where $\Delta x > 100$ mm and $\Delta y < 25$ mm were selected, thus limiting the effect of $\sigma_{\Delta y}\mbox{(stat)}$ on the overall error. \label{fig:R2xy}}
\end{center}
\end{figure}

In \cite{low_E_alphas04} the accuracy of an earlier SRIM version (SRIM2000) was found to be within $\sim 10$\%. In this work SRIM2003~\cite{SRIM} was used to predict alpha particle ranges in low pressure (40 Torr) CS$_2$ for comparison with the data. Since the mean angle between the tracks and the normal to the wireplane was $45^{\circ}$, the mean projected range on this plane, $\overline{R2_{xy}}$, was assumed to be equal to $\overline{\Delta z}$. The collimator allowed a flat distribution of angles between $(40 \pm 0.7)^{\circ}$ and $(50 \pm 0.5)^{\circ}$, where the estimated systematic errors arose from $\sim 1$ mm errors in the collimator dimensions. The mean angle of tracks through the collimator was $\theta = (45 \pm 0.18\ (\mbox{stat})\pm 1\ (\mbox{sys}))^{\circ}$, where the statistical error is that expected for a random sample of 241 tracks passing through the collimator and the systematic error was derived from errors in the collimator dimensions of $\pm 1$ mm. Based on these assumptions an independant confirmation of $\overline{\Delta z}$ was extracted from the data and compared with values derived from SRIM2003. Figure~\ref{fig:R2xy} shows the distribution of $R2_{xy}$ for tracks selected to have $\Delta x > 100$ mm and $\Delta y < 25$ mm, where:
\begin{equation}
\label{eq:R2xy-def}
R2_{xy} = \sqrt{(\Delta x)^2 + (\Delta y)^2}
\end{equation}

The mean of the values calculated using Equation~\ref{eq:R2xy-def} was determined as $\overline{R2_{xy}} = 127 \pm 1$ (stat) $\pm\ 4$ (sys) mm where the systematic error is based on the assumptions that $\sigma_{\Delta x}(\mbox{sys}) = 2$ mm and $\sigma_{\Delta y}(\mbox{sys}) \sim \overline{\Delta y}$. The effect of changing the value of the $\Delta y$ cut is shown in Table~\ref{tab:delta-y_cut} where the overall error is found to be close to a local minimum when the cut value is set to 25 mm. As the $\Delta y$ cut becomes more severe, $\Delta y$ and thus $\sigma_{\Delta y}(\mbox{sys})$ are reduced, however, the rapid drop in statistics causes an increase in $\sigma_{\Delta y}(\mbox{stat})$.

\begin{table}
\begin{center}
\caption{Variation in combined statistical and systematic errors on the value of $R2_{xy}$ as the maximum value of $\Delta y$ is changed. In all cases, the cut $\Delta x > 100$ mm was also imposed.\label{tab:delta-y_cut}}
\begin{tabular}{c|c|c|c|c}
$\Delta y$ cut & $\overline{R2_{xy}}$ & $\sigma_{R2}$ (stat) & $\sigma_{R2}$ (sys) & $\sigma_{R2}$ (combined)\\
\hline
15 mm & 130.9 mm & 6.5 mm & 1.8 mm & 6.7 mm\\
25 mm & 127.3 mm & 0.6 mm & 4.0 mm & 4.1 mm\\
35 mm & 124.4 mm & 0.5 mm & 5.9 mm & 5.9 mm\\
\end{tabular}
\end{center}
\end{table} 

With reference to Figure~\ref{fig:v-drift_from_alphas} the residual z-component of a track is given by:
\begin{equation}
\label{eq:Po210-dz-def}
\overline{\Delta z} = \overline{R2_{xy}} \cot (\overline{\theta}) = 127 \pm 1\ \mbox{(stat)} \pm 6\ \mbox{(sys) mm},
\end{equation}

where the error $\sigma_{\Delta z}$ on $\overline{\Delta z}$ is obtained via the usual law of error propagation taking into account the uncertainties in $\theta$ and $\overline{R2_{xy}}$.

The mean alpha particle range $\overline{R}$ is given by:
\begin{equation}
\label{eq:Po210-R-def}
\overline{R} = d \sec (\theta) + \overline{R2_{xy}} \csc (\theta) = 349 \pm 1\ \mbox{(stat)} \pm\ 6\ \mbox{(sys) mm},
\end{equation}

where $d = 120 \pm 1$(sys) mm is the distance from the source to the inner grid plane. The value for $\overline{R}$ should be compared to the SRIM2003 estimate of $R_{SRIM} = 317 \pm 13$ mm---suggesting a range correction-factor for typical alpha particle energies in 40 Torr CS$_2$ of $C_r = \overline{R}/R_{SRIM} = 1.10 \pm 0.05$, consistent with the $\sim 10$\% accuracy figure of \cite{low_E_alphas04}, mentioned above.

A total of 1410 alpha particle events were selected using the cuts described in \S~\ref{sec:flags}. This was reduced to 241 events by the $\Delta x$ and $\Delta y$ cuts mentioned above. The mean drift time was $\overline{\Delta t} = 2252 \pm 23 \mbox{ (stat)} \pm 10 \mbox{ (sys)} \mu$s, where the statistical error was derived from the spread in the data and the systematic error is an estimate of the constraint on timing accuracy due to the shaping time of the amplifiers. Combining with the measured value of $\overline{\Delta z}$ gives a drift velocity:
\begin{equation}
v_d = \overline{\Delta z}/\overline{\Delta t} = (57 \pm 1 \mbox{(stat)} \pm 3 \mbox{(sys)) ms$^{-1}$}.
\end{equation}

The only significant operational difference in DRIFT-IIa, compared to DRIFT-IIb was a slightly higher drift field (by a factor 1.067). Hence the drift velocity in DRIFT-IIa was estimated to be $v_d = (61 \pm 1 \mbox{(stat)} \pm 3 \mbox{(sys)})$ ms$^{-1}$.

\subsection{Background $^{222}$Rn Alpha Particles}
\label{sec:bg_alphas}
\begin{figure}
\begin{center}
\scalebox{0.7}{\includegraphics{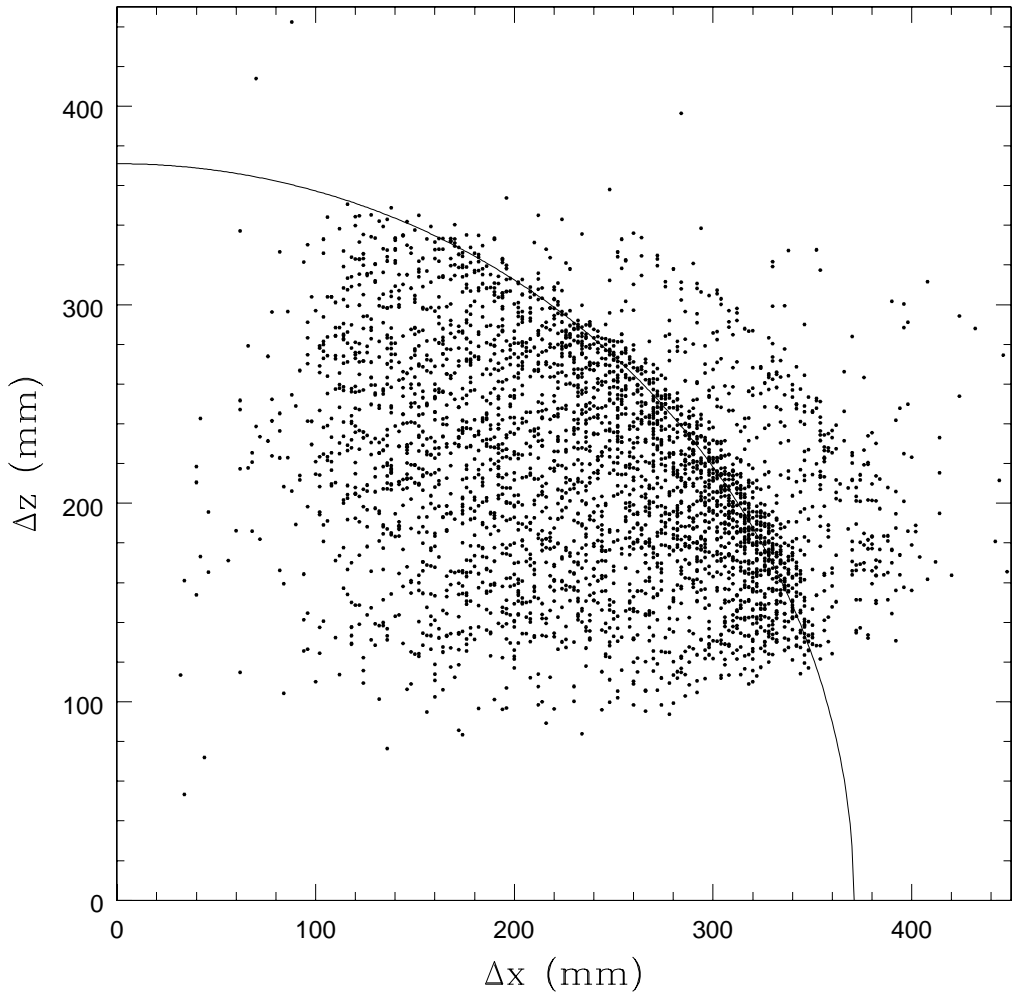}}
\caption{Example scatter plot of $x$ and $z$ components of GPCC tracks taken from a DRIFT-IIa dataset. A drift velocity of 59.2 ms$^{-1}$ was assumed to calculate $\Delta z$. A circular arc of radius 371 mm is shown, representing the maximum range of $^{222}$Rn alpha particles. Note that the higher range of $^{218}$Po alpha particle tracks is indicated by an arc-like feature at larger radius (see main text for details).\label{fig:IIa_x-y}}
\end{center}
\end{figure}

For this study, the intrinsic alpha particle background in DRIFT-IIa was used rather than a deliberately placed source. Under the assumption that these events, and in particular GPCC events (\S\ref{sec:flags}), were randomly selected from an isotropically distributed population, it was concluded that all possible track directions were equally represented in a suitably large sample. However, an inspection of the $x$ and $y$ components of tracks revealed a systematic underestimate of $\Delta y$, similar to the effect observed in DRIFT-IIb data (\S\ref{sec:Po210_alphas}), while the $x$ and $z$ components were found to have very similar distributions (see Figure~\ref{fig:IIa_x-y}). As in \S\ref{sec:Po210_alphas}, this effect was found to reduce the apparent $\Delta y$ measurement by a factor of two, implying a sytematic error $\sigma_{\Delta y} \sim \Delta y$. To reduce the influence of this error on the full 3D track range a cut was applied that excluded tracks with $\Delta y > 60$ mm. An attempt could then be made to discriminate on range to select a single track species (for example, alpha particle tracks from $^{222}$Rn decay) from the data. As such, the total number of events was reduced from 3498 to 1031.

\begin{figure}
\begin{center}
\scalebox{0.7}{\includegraphics{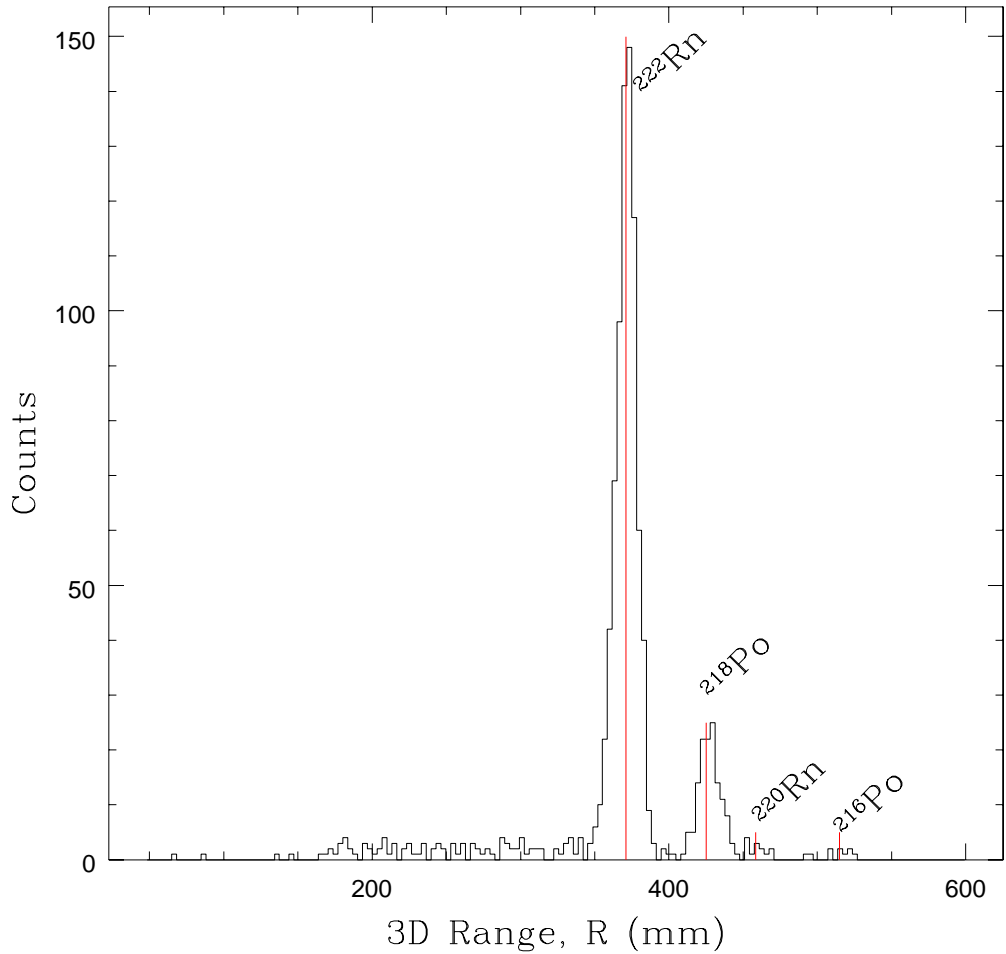}}
\caption{Histogram of the 3D range $R$ for GPCC alpha events in DRIFT-IIa selected to have $\Delta y < 60$ mm. The peaks are attributed to alpha decays of $^{222}$Rn (at $R \simeq 371 $ mm), $^{218}$Po ($R \simeq 425$ mm), $^{220}$Rn ($R \simeq 458$ mm) and $^{216}$Po ($R \simeq 515$ mm). Vertical lines indicate the mean peak positions. These ranges are $10\%$ higher than those appearing in Table~\ref{tab:alpha_ranges}. A low-level continuum is also visible below $R \sim 350$ mm, attributed to alpha particles that traversed part of a cathode wire and subsequently escaped into the gas with degraded energy.\label{fig:R3_discrim}}
\end{center}
\end{figure}

Analysis of the GPCC events revealed that alpha particle tracks arising from the decay of four isotopes could be separately identified. Figure~\ref{fig:R3_discrim} shows a histogram of track range $R$ for GPCC events from a DRIFT-IIa dataset where tracks were selected to have $\Delta y < 60$ mm. The range was calculated according to:

\begin{equation}
R = \sqrt{ (\Delta x)^2 + (\Delta y)^2 + (\Delta t \times v_d)^2 },
\end{equation}

where the drift velocity $v_d = 59.2$ ms$^{-1}$ was chosen to produce the narrowest peaks and therefore the best separation in the distribution shown in Figure~\ref{fig:R3_discrim}. The dominant peak in this case (within a region of interest (ROI) $\{ 345 \mbox{ mm} < R < 395 \mbox{ mm} \}$) was attributed to $^{222}$Rn. The mean value of the track range in this region (containing 768 events) was $R = 371.1 \pm 0.3$ (stat) $\pm\ 11$ (sys) mm (the systematic error was calculated under the assumption of a $\pm 1$ ms$^{-1}$ error in the assumed drift velocity). This range is larger than the SRIM-predicted track range listed in Table~\ref{tab:alpha_ranges} by $\sim 10$\%, consistent with the results in the previous section.

\begin{figure}
\begin{center}
\scalebox{0.7}{\includegraphics{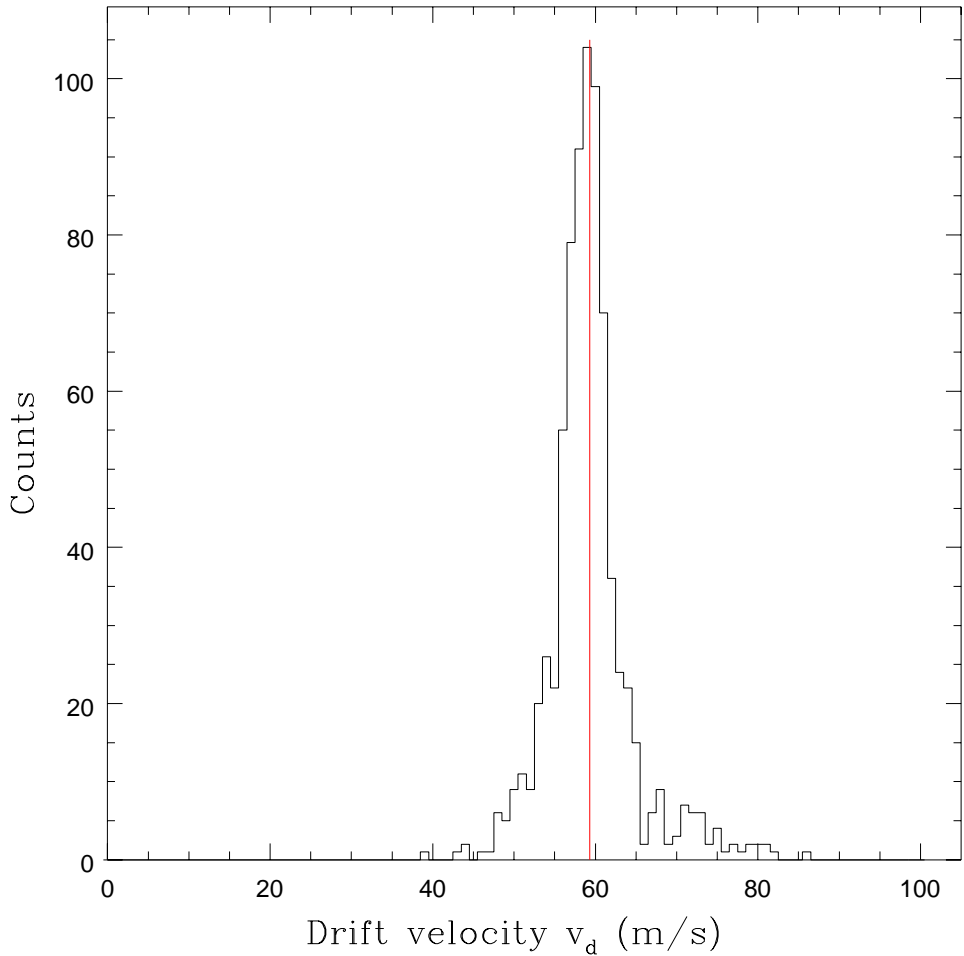}}
\caption{Distribution of drift-velocity measurements for alpha particle tracks from $^{222}$Rn decay in DRIFT-IIa selected to have $\Delta y < 60$ mm.\label{fig:v_D222-distr}}
\end{center}
\end{figure}

The above range discrimination procedure was used as a cut to preferentially select $^{222}$Rn decay events from the GPCC population, since these were the most numerous and thus subject to the lowest level of contamination from other decay events. Specifically, tracks were selected with a dominant $x$-component by selecting only those tracks with $\Delta y < 60$ mm to reduce the influence of underestimated $\Delta y$ values. For each $^{222}$Rn event, the $x$ and $y$ components, $\Delta x$ and $\Delta y$ and deposition time $\Delta t$ were measured with assumed systematic errors of $\sigma_x = 2$ mm, $\sigma_y \sim \Delta y$, $\sigma_t = 10\ \mu$s. In addition, tracks were then selected that fell within the range region of interest  $\{ 345 \mbox{ mm} < R < 395 \mbox{ mm} \}$, calculated for the case where drift velocities were in the range $\{57\mbox{ ms$^{-1}$} < v_d < 63\mbox{ ms$^{-1}$} \}$ and coinciding with the dominant peak. The drift velocity $v_D$ for each track was then calculated according to:
\begin{equation}
\label{eq:v_D_from_Rn-alphas}
v_D = \frac{ \sqrt{R_{222}^2 - \Delta x^2 - \Delta y^2} }{\Delta t},
\end{equation}
where the $^{222}$Rn track range $R_{222}$ was treated as a free parameter and varied until the sampling error of the resulting distribution of $v_D$ tended to a minimum value. Using this procedure, the best estimate of the drift velocity in DRIFT-IIa, derived from $^{222}$Rn alpha tracks, was $59.3 \pm 0.2$ (stat) $\pm 7.5$ (sys) ms$^{-1}$ when $R_{222} = 371$ mm.  The distribution of $v_d$ in DRIFT-IIa, determined with the best fit value of $R_{222}$, is shown in Figure~\ref{fig:v_D222-distr}. For DRIFT-IIb, the corresponding value was estimated to be $v_D = 55.6 \pm 0.2$ (stat) $\pm 7$ (sys) ms$^{-1}$. These values are consistent with earlier results.

There are several useful pieces of information provided by the plot in Figure~\ref{fig:R3_discrim};
\begin{enumerate}
\item{The relative peak positions, in order of increasing $R$, match very closely with those expected for the alpha decays of $^{222}$Rn, $^{218}$Po, $^{220}$Rn and $^{216}$Po if a correction factor of $\sim 10\%$ is applied (see Table~\ref{tab:alpha_ranges}).}
\item{The fact that cathode-crossing alpha tracks are observed with a range consistent with $^{218}$Po (and possibly $^{216}$Po) suggests that not all radon progeny plate out---some progeny remain suspended in the gas for long enough to decay and produce GPCC events. The most likely explanation for this is that some of the progeny are produced \emph{uncharged}.}
\item{On this assumption, since each $^{222}$Rn decay results in exactly one $^{218}$Po decay, the ratio of the statistics in the two main peaks, adjusted for their relative observation efficiencies, determines what fraction $f_U$ of $^{218}$Po progeny are produced uncharged in 40 Torr CS$_2$.}
\item{A similar calculation performed on the next two peaks can be used to determine the uncharged fraction of $^{216}$Po in 40 Torr CS$_2$.}
\end{enumerate}
The determination of the values of $f_U$, as suggested by points (3) and (4) above, is discussed in \S\ref{sec:MCs}.

Table~\ref{tab:v_drift-results} summarises all the drift velocity measurements obtained in this section. It should be noted that the assumption $\sigma_{\Delta y} \mbox{ (sys)} \sim \Delta y$ yields a conservative estimate for the systematic uncertainty since the change in track range (and hence in $v_D$) with the $\Delta y$ cut is smaller than this uncertainty (see Table~\ref{tab:delta-y_cut}).

\begin{table}
\begin{center}
\caption{Results of drift velocity measurements. Values in [brackets] are derived values based on the drift field ratio between the two modules.\label{tab:v_drift-results}}
\vspace{2mm}
\begin{tabular}{c|c|c}
Method                          & $v_D$ [ms$^{-1}$], DRIFT-IIa & $v_D$ [ms$^{-1}$], DRIFT-IIb\\
\hline
Collimated $^{210}$Po $\alpha$s & [$61 \pm 1 \mbox{ (stat)} \pm 3 \mbox{ (sys)}$] & $56 \pm 1.7 \mbox{ (stat)} \pm 3 \mbox{ (sys)}$ \\
$^{222}$Rn $\alpha$ background  & $59.3 \pm 0.2 \mbox{ (stat)} \pm 7.5 \mbox{ (sys)}$ & [$55.6 \pm 0.2 \mbox{ (stat)} \pm 7 \mbox{ (sys)}$]\\
\end{tabular}
\end{center}
\end{table}

The track range measurements show a small deviation from SRIM2003 range estimates for alpha particles in 40 Torr CS$_2$. A. Mangiarotti et al.~\cite{Mangiarotti07} note inaccuracies in SRIM for heavy ions, mainly due to discrepancies in the electronic contribution to the energy loss. Electronic stopping is by far the most dominant energy-loss mechanism for alpha particles in CS$_2$. The authors of SRIM quote a mean deviation from experimental data of $\sim 4.1\%$ for the stopping power of He ions\cite{SRIM}.

\section{Monte Carlo Simulation of Alpha Activity}
\label{sec:MCs}
A Monte Carlo code was written to determine the total rate of radon decays $D_{Rn}$ occuring in the detector given an observed rate $g\times D_{Rn}$ of GPCCs, where $g$ is a geometric factor dictated by the cuts outlined in \S\ref{sec:flags} equal to the proportion of events that would be observed as valid GPCCs. From this, and a knowledge of the gas flow-rate $F$ and total mass of gas $m_o$ in the detector, the rate $E_{Rn} = k/g$ of radon emanation occuring within the vessel could be determined, where $k$ is the rate of emanated radon atoms emitting an alpha particle that is then observed as a GPCC event. These quantities, for a given radon species, are related by the expression:

\begin{equation}
\label{eq:Sean_eq}
D_{Rn} = \frac{k}{g\left(1+\tau_e/\rho \right)} \left(1 - \exp\left[\frac{-t}{\tau_e \rule[-1mm]{0mm}{2mm}} \right]\times \exp\left[\frac{-t}{\rho} \right]  \right),
\end{equation}

where $t$ is the elapsed time from the point at which radon starts to emanate into the vessel after initial evacuation and flushing with uncontaminated gas, $\tau_e = \tau_{1/2}/\ln(2)$ is the characteristic decay time of the radon (where $\tau_{1/2}$ is the half-life) and $\rho$ is the characteristic flush time of the vessel, related to the flow rate by:

\begin{equation}
\label{eq:flush-time}
\rho = \frac{m_0}{F}\left(1-1/e\right)
\end{equation} 

The typical flow rate for DRIFT-IIa operation was $F = 0.118$ kg/day and the corresponding flush-time was $\rho = 3.15$ days. At a pressure of 40 Torr $m_0 = 588$ grammes for the $1.5 \times 1.5 \times 1.5$ m$^3$ vessel plus input pipework.

The simulation modelled alpha particle tracks due to the decay of $^{222}$Rn, $^{218}$Po, $^{220}$Rn or $^{216}$Po. The process was simplified by assuming all alpha particle tracks were straight and had ranges $C=1.1$ times greater than those listed in Table~\ref{tab:alpha_ranges}. The range correction factor $C$ was applied to reflect the slightly longer ranges observed in the data compared to SRIM2003 predictions (see \S\ref{sec:Po210_alphas} and \S\ref{sec:bg_alphas}).

Tracks for each of the decays were simulated in separate runs to determine (i) the value of the geometric factor $g$ for radon decay events for a given drift velocity and range correction factor $C$, (ii) the effect of varying the drift velocity, $C$ and polonium uncharged fraction $f_U$ on the relative numbers of radon and polonium GPCCs. The best matches to the observed ratios of event counts seen in Figure~\ref{fig:R3_discrim} were found when $f_U (^{218}\mbox{Po}) = (22 \pm 2)$\% and $f_U (^{216}\mbox{Po}) = (100^{+0}_{-35})$\%.

All tracks were defined in terms of a start point and end-point where the end point was chosen to lie in a random direction but at a fixed distance, specified by the alpha particle range, from the start point. Radon start points were randomly chosen to lie anywhere within the vessel volume, as were uncharged polonium start points. Charged polonium start points were randomly picked from a one m$^2$ plane coinciding with the central cathode. Table~\ref{tab:fidMC_results} lists the outputs from typical runs in which the uncharged fractions for $^{218}$Po and $^{216}$Po were set at 28\% and 100\%, respectively. These $f_U$ values were chosen to produce the best match to observed ratios of radon to polonium progeny (see \S\ref{sec:bg_alphas} and Figure~\ref{fig:R3_discrim}). The value of $g$ is given by the ratio of the number in line VI of the table to the total number of simulated alpha decays ($10^5$).

\begin{table}
\begin{center}
\caption{Results from simulation of $10^5$ alpha decays. The entries are the numbers of events that satisfied the following criteria: I. Track confined within vessel; II. At least partly in fiducial volume (FV); III. Fully within FV; IV. In FV and also crossing central cathode; V. In FV and $z < z_{max}$; VI. In FV, $z < z_{max}$, crossing central cathode and $>8$ anode hits on each side (GPCC); VII. GPCC with $\Delta y < 120$ mm (see \S\ref{sec:bg_alphas} for the motivation for this constraint).\label{tab:fidMC_results}}
\begin{tabular}{c|l|c|c|c|c}
$v_D$ (ms$^{-1}$) & Track category & $^{222}$Rn & $^{218}$Po & $^{220}$Rn & $^{216}$Po\\
\hline
59.8    &I       & 66652      & 84327      & 73093      & 55665\\
(DRIFT-IIa) &II  & 27847      & 77736      & 70679      & 35852\\
        &III     & 15581      & 48084      & 27889      & 11401\\
        &IV      & 3877       & 1159       & 673        & 4577\\
        &V       & 8360       & 17822      & 8413       & 4535\\
        &VI      & 1583       & 291        & 83         & 740\\
        &VII     & 197        & 33         & 3          & 52\\
\hline
56.1    &I       & 66751      & 84309      & 73152      & 55643\\
(DRIFT-IIb) &II  & 27900      & 77714      & 70725      & 35931\\
        &III     & 15602      & 48052      & 27794      & 11438\\
        &IV      & 3876       & 1176       & 651        & 4616\\
        &V       & 7867       & 16611      & 7856       & 4226\\
        &VI      & 1161       & 261        & 63         & 645\\
        &VII     & 160        & 25         & 3          & 44\\
\end{tabular}
\end{center}
\end{table}

The values of the geometric factors in DRIFT-IIa for both $^{222}$Rn and $^{220}$Rn, determined from multiple runs of the code, were $g = (13.64 \pm 0.16\mbox{ (stat) }\pm 0.60\mbox{ (sys)})\times 10^{-3}$ and $g = (9.63 \pm 0.09\mbox{ (stat) }\pm 0.57\mbox{ (sys)})\times 10^{-3}$, respectively, due to the cuts described in~\S\ref{sec:flags}. The difference is due to the lower probability of the longer $^{220}$Rn alpha particle tracks being fully contained within the effective fiducial volume. A similar difference is seen in DRIFT-IIb ($g = (12.18 \pm 0.11\mbox{ (stat) }\pm 0.55\mbox{ (sys)})\times 10^{-3}$ for $^{222}$Rn, $g = (8.62 \pm 0.09\mbox{ (stat) }\pm 0.45\mbox{ (sys)})\times 10^{-3}$ for $^{220}$Rn) where the lower drift field reduced the fiducial volume in the z-direction. In all cases the systematic error takes into account uncertainties in the track ranges and drift velocities and the statistical error represents the spread in the results from ten identical runs.

\begin{figure}
\begin{center}
\scalebox{0.7}{\includegraphics{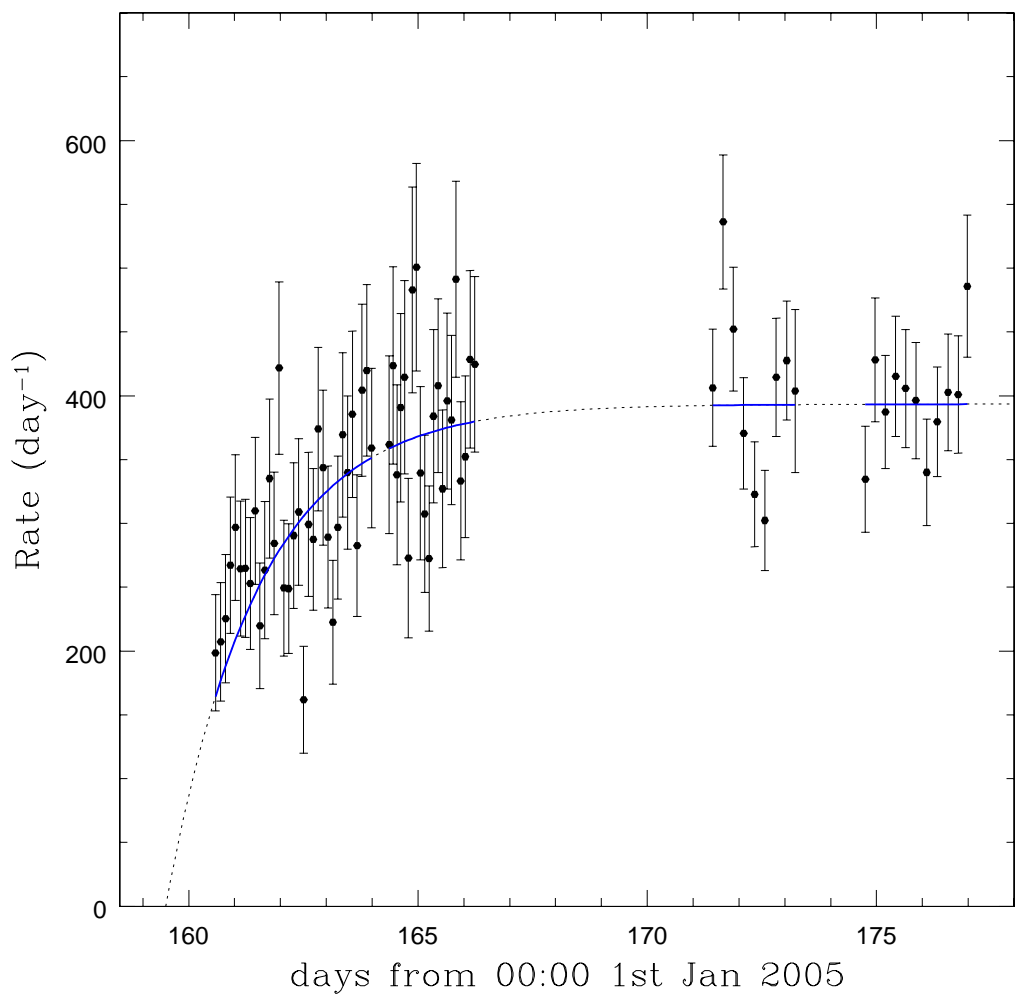}}
\caption{Rate of GPCC events observed in DRIFT-IIa. Also shown is a fit to the data using the function given in Equation~\ref{eq:Sean_eq}.\label{fig:alpha-rate_fit}}
\end{center}
\end{figure}

The rate of GPCC events observed in DRIFT-II background data was used to estimate the total emanation rate into the vessel from all components. Figure~\ref{fig:alpha-rate_fit} shows data taken from a series of DRIFT-IIa runs spanning a period of about two and a half weeks in the middle of the 2005 data taking period. These particular runs were chosen because the vessel had been evacuated, flushed and refilled immediately before acquisition started. A fit to the function given in Equation~\ref{eq:Sean_eq} is also shown where the only free parameter was the reduced emanation rate $k$. A clear exponential rise is observed in accordance with the model described by Equation~\ref{eq:Sean_eq} with a best fit value of $k = (13.6 \pm 1.5) \times 10^{-3}$ Hz.

Setting $t = \infty$ in Equation~\ref{eq:Sean_eq} allows one to determine the steady-state radon decay rate as $D_{Rn} = 0.4 \pm 0.04$ Hz and thus the total steady-state emanation rate was $E_{Rn} = 1.0 \pm 0.1$ Hz, assuming $^{222}$Rn formed the dominant contribution to the observed alpha activity.

\section{Conclusions}
\label{sec:conclusions}
An independant analysis, separate from that used to search for nuclear recoil events, has been developed to study phenomena associated with the alpha decays of radioactive contamination within the DRIFT-IIa and DRIFT-IIb directional dark matter detectors.

The alpha particles produced as a result of the decay of $^{222}$Rn, $^{220}$Rn and their progeny (easily rejected in a dark matter analysis) have proved useful in determining a number of operational characteristics. The drift velocity was determined by two methods, one in each of the operational modules (see Table~\ref{tab:v_drift-results} in previous section). The first method, employing a directed $^{210}$Po alpha particle source in DRIFT-IIb determined the drift velocity to be $v_D = 56 \pm 1.7$ (stat) $\pm 3$ (sys) ms$^{-1}$. The second method, using the intrinsic alpha particle background in DRIFT-IIa found $v_D = 59.3 \pm 0.2$ (stat) $\pm 7.5$ (sys) ms$^{-1}$. Additionally, the range of alpha particle tracks was found to be an efficient discriminating factor between different alpha particle emitting species. This highlighted the possibility of some radon progeny being produced within the fiducial volume in an uncharged state and thus not being swept onto the central cathode by the drift field. Consequently, the decay of these uncharged progeny were able to produce alpha particle tracks that occasionally cross the cathode (GPCC events). Combining knowledge of the relative populations of radon GPCCs and those of their immediate progeny with results of a Monte Carlo simulation allowed an estimate of the fraction $f_U$ of polonium progeny produced uncharged. In the case of $^{218}$Po, the uncharged fraction was found to be $f_U = 22 \pm 2 \%$ and for $^{216}$Po, $f_U = 100^{+0}_{-35} \%$. This knowledge may prove useful in future more complex simulations of alpha emitting contamination in DRIFT-type detectors.

\section{Acknowledgements}
\label{acknowledgements}
We aknowledge the support of the US National Science Foundation (NSF). This material is based upon work supported by the National Science Foundation under Grant Numbers 0300973 and 0600789. Any opinions, findings and conclusions or recommendations expressed in this material are those of the author(s) and do not necessarily reflect the views of the National Science Foundation. We also acknowledge the financial support from the EU FP6 programme ILIAS (Contract RII3-CT-2004-506222), PPARC and the New Mexico Center for Particle Physics. CG and SJSP are grateful to EPSRC for the support of their PhD research. The Collaboration would also like to thank the staff of Cleveland Potash Ltd. for their assistance.

\end{document}